\documentclass[usenatbib,usegraphicx,useAMS,letters]{mn2e}
\usepackage{amsmath}
\usepackage{amssymb}

\def\eg{{e.g.}} 
\def\ie{{i.e.}} 
\renewcommand{\d}{{\rm d}}

\title{Radiative Transfer Along Rays in Curved Spacetimes}
\author[Avery E. Broderick]{Avery E. Broderick\thanks{E-mail:
    abroderick@cfa.harvard.edu}\\ Institute for Theory and 
  Computation, Harvard-Smithsonian Center for Astrophysics, 60 Garden
  St., MS 51, Cambridge, MA 02138, USA\\}

\begin{document}
\maketitle

\begin{abstract}
Radiative transfer in curved spacetimes has become increasingly
important to understanding high-energy astrophysical phenomena and
testing general relativity in the strong field limit.  The equations
of radiative transfer are physically equivalent to the Boltzmann
equation, where the latter has the virtue of being covariant.  We show
that by a judicious choice of the basis of the phase space, it is
generally possible to make the momentum derivatives in the Boltzmann
equation vanish along an arbitrary (including nongeodesic) path, thus
reducing the problem of radiative transfer along a ray to a path
integral in coordinate space.
\end{abstract}


\section{Introduction} \label{I}
Radiative transfer in general relativistic environments is required to
understand many high-energy astrophysical phenomena.  Despite
this, the approaches utilised are either ad hoc, use complicated
transfer functions \citep[see, \eg,][]{Cunn:75}, moment formalisms
\citep[see, \eg,][]{Thor:81}, Monte-Carlo methods
\citep[see, \eg,][]{Jank-Hill:89}, or the Boltzmann equation
is solved in its entirety
\citep[see, \eg,][and references therein]{Lind:66,Lieb-Ramp-Jank-Mezz:05}.
Many researchers who are interested in producing images or light
curves of compact objects have tacitly assumed that the problem of
radiative transfer is reduced to performing an integral along the
line-of-sight \citep[see, \eg,][]{Jaro-Kurp:97, Brod-Blan:04}.
While being intuitively obvious (radiative transfer is a local
process), we have been unable to find an explicit justification for
this in the literature.  Here we show that by Fermi-Walker propagating
a basis along the ray the Boltzmann equation may indeed be reduced to
a path integral along the ray (as long as scattering is ignored).

\section{Boltzmann Equation}
The Boltzmann equation is given by
\begin{equation}
\left.
\frac{\d x^\mu}{\d\tau} \frac{\partial \mathcal{N}}{\partial x^\mu}
\right|_{p^\mu}
+
\left.
\frac{\d p^\mu}{\d\tau} \frac{\partial \mathcal{N}}{\partial p^\mu}
\right|_{x^\mu}
=
\mathcal{S}(x^\mu,p^\mu,\mathcal{N})\,,
\label{bz1}
\end{equation}
where $\tau$ is an affine parameterisation of the ray, $p^\mu$ is the
tangent to the ray, $\mathcal{N}$ is the distribution function of the particles
under consideration, and $\mathcal{S}$ contains the source
terms \citep{Lind:66}\footnote{In general, it is also possible to write this
in terms of the conjugate coordinate-momentum pair $x^\mu$ and
$p_\mu$, with identical results.}.  The particular form of
$\mathcal{S}$ and that it is a Lorentz scalar are discussed in
\citet{Lind:66} and \citet{Brod-Blan:04}.  As shown in
\citet{Lind:66}, equation (\ref{bz1}) can be written in terms of an
arbitrary tangent-space basis ($\{e_a^\mu\}$) as
\begin{equation}
\left.
\frac{\d x^\mu}{\d\tau} \frac{\partial \mathcal{N}}{\partial x^\mu}
\right|_{p_a}
+
\left.
\frac{\d p_a}{\d\tau} \frac{\partial \mathcal{N}}{\partial p_a}
\right|_{x^\mu}
=
\mathcal{S}(x^\mu,p_a,\mathcal{N})\,,
\end{equation}
where  $p_a \equiv p_\mu e_a^\mu$.
The momentum terms will generally vanish if $\d p_a/\d\tau = 0$, which is
true by definition if $\{e_a^\mu\}$ are Fermi-Walker
transported\footnote{Fermi-Walker transport is defined by
\begin{equation*}
u^\nu \nabla_\nu f^\mu
=
\left( u^{\mu} a^{\nu} - a^{\mu} u^{\nu} \right) f_\nu \,,
\end{equation*}
where $u^\mu$ is the unit tangent vector ($u^\nu u_\nu = -1$) of the
ray and $a^\mu\equiv u^\nu\nabla_\nu u^\mu$.  If the ray is a geodesic
$a^\mu=0$ and this trivially reduces to parallel transport.}
along the ray \citep[see, \eg,][\S 6.5]{Misn-Thor-Whee:73}, resulting in
\begin{equation}
\left.
\frac{\d x^\mu}{\d\tau} \frac{\partial \mathcal{N}}{\partial x^\mu}
\right|_{p_a}
=
\mathcal{S}(x^\mu,p_a,\mathcal{N})\,.
\label{bz3}
\end{equation}
If the path is a geodesic, Fermi-Walker transport simply reduces to
parallel transport.  It is worth noting that the condition
$\d p_a/\d\tau=0$ will also be satisfied if $e_a^\mu$ is a
Killing vector.

\section{Example Applications}
The formulation of radiative transfer as a path integral along the ray
in coordinate space (as opposed to in the entire phase space) is
especially convenient, \eg, for modeling high-resolution images and
light curves of compact objects
\citep[see, \eg,][]{Brod-Loeb:05b,Brod-Loeb:05,Brod-Loeb:05c} and
ray-casting radiative transfer codes
\citep[see, \eg,][]{Razo-Scot:99,Soka-Abel-Hern:01}.  In the former,
typically a congruence of rays are traced from a distant observer
backwards in time towards an emitting region.  In the latter an
isotropic distribution of rays are traced outward from each grid
point.  Using the radiative transfer equation in the form of
equation (\ref{bz1}) generally requires that the complete distribution
function be constructed at each point along each ray.  In contrast,
equation (\ref{bz3}) is trivially integrated along the ray at a single
observing frequency (which for imaging purposes is usually the
situation of interest).  This is true even when the rays are not
geodesics, \eg, in the presence of strong refraction
\citep[see, \eg][]{Brod-Blan:03}.

For vacuum photon propagation in the Kerr spacetime (that which is
most likely to be astrophysically relevant), and in Petrov type-D
spacetimes more generally, the transported basis $\{e_a^\mu\}$ may be
constructed algebraically.  This is done by first noting that the
tangent vector to the ray is already parallely propagated by definition, and thus
choose $e_0^\mu \propto \d x^\mu/\d \tau$.  Two fiducial vectors,
$e_1^\mu$ and $e_2^\mu$, orthogonal to each other and $e_0^\mu$, are
then chosen at some position along the ray.  These may be transported
along the ray algebraically by using the orthogonality relations, a
normalization condition, and the complex constant first
described by \citet{Walk-Penr:70}
\citep[though see ][ as well]{Chan:92}\footnote{For the Kerr
  spacetime, in Boyer-Lindquist coordinates the Penrose-Walker
  constant is given by (\eg, for $e_1^\mu$)
\begin{multline}
\kappa_{\rm PW} =\\
\left( r - i a \cos\theta \right)
\left\{ \left( e_0^t e_1^r - e_0^r e_1^t \right)
+ a \left( e_0^r e_1^\phi - e_0^\phi e_1^r \right) \sin^2 \theta
\right.\\
\left. - i \left[ \left(r^2 + a^2\right) \left(e_0^\phi e_1^\theta -
  e_0^\theta e_1^\phi \right)
- a \left( e_0^t e_1^\theta - e_0^\theta e_1^t \right) \right]
\sin\theta \right\}\,.
\end{multline}
}.
Finally, the fourth basis-vector is then obtained from the conditions
\begin{equation}
e_4^\mu e_{0\,\mu}^{\phantom{\mu}} = 1\,,
\quad
e_4^\mu e_{4\,\mu}^{\phantom{\mu}} = e_4^\mu e_{1\,\mu}^{\phantom{\mu}} = e_4^\mu e_{2\,\mu}^{\phantom{\mu}} = 0\,.
\end{equation}
If desired, an orthonormal basis can be created from $\{e_a^\mu\}$ by
making a linear combinations of $e_0^\mu$ and $e_4^\mu$.

However, for many applications, the explicit construction of
$\{e_a^\mu\}$ is unnecessary.  For example, consider the case when the
source terms $\mathcal{S}$ depend only upon $\mathcal{N}$, position
($x^\mu(\tau)$) and the momentum of photons moving {\em along} the ray
($\wp^\mu(\tau)\propto\d x^\mu/\d\tau$), \ie, anisotropic scattering
into and out of the ray may be neglected.  Thus, only vector
$e_0^\mu$, as defined above, is required. However, this is presumably
already available from the ray construction.

\begin{figure}
\begin{center}
\includegraphics[width=\columnwidth]{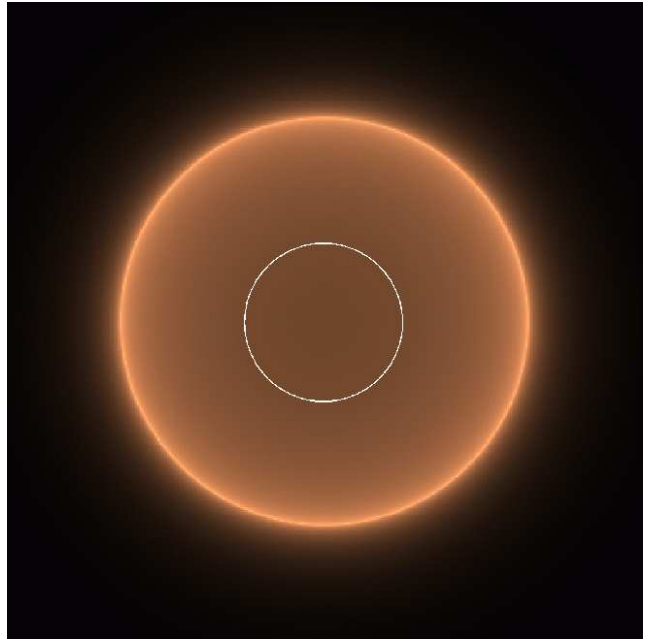}
\end{center}
\caption{The intensity map of a stationary, non-thermal, self-absorbed
  synchrotron emitting plasma surrounding a non-rotating black hole as
  seen at infinity for a single frequency.  The contrast is linear
  ranging from maximum intensity to vanishing intensity (black).  The
  density ($\propto r^{-3/2}$) and magnetic field strength ($\propto
  r^{-5/4}$) are spherically symmetric and the latter is assumed to
  be randomly oriented (\ie, the emission and absorption have been
  averaged over field orientations).  The spectral index of the
  emitting electrons is taken to be 2.  For reference, the white ring
  shows the size of the horizon.}
\label{fig1}
\end{figure}

An astrophysically example in which this is the case is self-absorbed
synchrotron emission.
In this case $\mathcal{S} = j - a \mathcal{N}$ where
for a power-law electron distribution,
\begin{equation}
\begin{aligned}
j &= j_0 \frac{n_e}{\omega} \left(\frac{\omega_B \sin\vartheta}{\omega}\right)^{\alpha+1}\\
a &= a_0 n_e \left( \frac{\omega_B\sin\vartheta}{\omega} \right)^{\alpha+3/2}\,,
\end{aligned}
\end{equation}
where $j_0$ and $a_0$ are constants, $\alpha$ is
the spectral index, $n_e$ is the proper electron number density,
$\omega$ and $\omega_B$ are the photon and cyclotron frequency in the
plasma rest frame, respectively, and $\vartheta$ is the angle between
the magnetic field and the ray in the plasma rest frame.  Clearly,
along the ray $n_e$ and $\omega_B$ are functions of $x^\mu(\tau)$
only.  The photon frequency as measured in the rest frame is given by
$u^\mu \wp_\mu$ (where $u^\mu$ is the plasma four-velocity) and thus
is a function of $x^\mu(\tau)$ and $\wp^\mu(\tau)$ only.  Finally, the angle $\vartheta$ may be defined by
its cosine, which is in turn given by $\wp^\mu u^\nu
\mbox{}^*\!F_{\mu\nu} / \omega \omega_B$ where $\mbox{}^*\!F^{\mu\nu}$
is the dual of the electromagnetic field tensor.  Thus, as claimed
$\mathcal{S}$ is a function soley of $\mathcal{N}(\tau)$,
$x^\mu(\tau)$ and $\wp^\mu(\tau)$.  This may then be inserted into
equation (\ref{bz3}), which can now be trivially integrated to yield
$\mathcal{N}(\wp^\mu_0)$, where $\wp^\mu_0$ enters via the initial
conditions of the ray (presumably at $\tau=0$).  An image produced by
this procedure is shown in Figure \ref{fig1}, and additional more
complex examples can be found elsewhere in the literature
\citep[see, \eg,][]{Brod-Blan:03,Brod-Blan:04,Brod-Loeb:05b,Brod-Loeb:05,Brod-Loeb:05c,Falc-Meli-Agol:00,Jaro-Kurp:97}.

\section{Discussion}

The simplification of the Boltzmann equation has a straightforward
physical interpretation.  Since radiative transfer is a local process,
curvature enters only in relating the tangent space at different
points of the spacetime.  For the observer moving along a
ray, by definition, this is naturally accounted for by Fermi-Walker
transporting a tangent space basis, thus reducing the problem to its
flat space analogue (for which the momentum derivative terms can
always be made to vanish).

While it is possible to define the basis $\{e_a^\mu\}$ on any three
dimensional hypersurface of the spacetime, unfortunately, it is not
generally possible to produce a well defined basis in the full spacetime as
a result of ray intersections.  This does not mean, of course, that it
is not possible to integrate the Boltzmann equation along any single
ray.  Rather, the phase space bases for different rays which
intersect will in general produce a different basis at the point of
crossing, and thus cannot be compared directly (though they may be
rotated).

\section*{Acknowledgements}
A.E.B would like to thank Jon McKinney and George Rybicki
for useful conversations.  This work was supported by an ITC
Fellowship from Harvard College Observatory.

\bibliographystyle{mn2e.bst}
\bibliography{grrt.bib}

\bsp

\end{document}